# Nonparametric species richness estimation under convexity constraint

Cécile Durot[1] `cecile.durot@u-paris10.fr`
Sylvie Huet[2] `sylvie.huet@jouy.inra.fr`
François Koladjo[2,3] `francois.koladjo@gmail.com`
Stéphane Robin[4,5] `stephane.robin@agroparistech.fr`

[1] UFR SEGMI, Université Paris Ouest Nanterre La Défense, F-92001, Nanterre, France

[2] UR341 MIA, INRA, F-78350 Jouy-en-Josas, France

[3] CIPMA-Chaire UNESCO, FAST, UAC, 072BP50 Cotonou, Bénin

[4] INRA, UMR518 MIA, AgroParisTech F-75005 Paris, France

[5] AgroParisTech, UMR518 MIA, AgroParisTech F-75005 Paris, France

**Abstract.** We consider the estimation of the total number $N$ of species based on the abundances of species that have been observed. We adopt a non parametric approach where the true abundance distribution $p$ is only supposed to be convex. From this assumption, we propose a definition for convex abundance distributions. We use a least-squares estimate of the truncated version of $p$ under the convexity constraint. We deduce two estimators of the total number of species, the asymptotic distribution of which are derived. We propose three different procedures, including a bootstrap one, to obtain a confidence interval for $N$. The performances of the estimators are assessed in a simulation study and compared with competitors. The proposed method is illustrated on several examples.

# 1 Introduction

Estimation of abundance is one of the oldest way to evaluate the diversity of species in a given area. The problem traces back to Fisher et al. (1943), who first proposed to estimate the distribution of abundance in a Gamma-Poisson framework. Several approaches have been considered and various sampling theoretic frameworks may be considered for modelling observations of species abundance in a population, see for example the review given by Bunge and Fitzpatrick (1993) or more recently by Bunge et al. (2014) in the context of microbial diversity estimation.

The generic problem can be stated as follows. Considering a population composed of $N$ species, for each $i = 1, \ldots, N$, let us denote by $A_i$ the abundance (that is, the number of observed individuals) of species $i$ in a sample of size $n$, and by $S_j$ the number of species having abundance $j$:

$$S_j = \sum_{i=1}^{N} I(A_i = j), \; j \in \{0, \ldots, n\}. \tag{1}$$

The connection between $N$ and the $S_j$'s is that $N = S_0 + D$, where

$$D = \sum_{j \geqslant 1} S_j = \sum_{i=1}^{N} I(A_i \geqslant 1) \tag{2}$$

denotes the number of observed species. Here, $S_0$ denotes the number of species that are indeed present in the population but have not been observed in the sample, so that $S_0$ is not observable whereas $S_j$ is observable for all $j \geqslant 1$, and $D$ also is observable. The variables $A_i$ are not completely observed: one observes only zero-truncated counts, which means that one observes only the $A_i$'s that are strictly positive. The problem is to estimate $N$ based on the observations $S_j, j \geqslant 1$, or equivalently, based on the observation of the zero-truncated counts. We briefly describe below the main approaches of modeling that have been investigated in the literature.

A first approach consists in considering that $n$ individuals are sampled from an infinite population composed of $N$ species in proportions $q_1, \ldots, q_N$. In the case of sampling with replacement, the vector $(A_1, \ldots, A_N)$ has a multinomial distribution with parameters $n$ and $q_1, \ldots, q_N$. In this setting, Harris (1959) considered the problem of estimating the sample coverage $\sum_i q_i I(A_i \geqslant 1)$ and predicting the number of observed species in enlarged samples. Moreover, he provided an approximation for the expected number of unobserved species $E(S_0)$. Inspired by this approximation, Chao (1984) proposed an estimator of a lower bound for $N$ without any assumption on the $q_i$'s. Chao illustrated on some examples that her estimator can be considered as an estimator for $N$ if $n$ is large and most of the information is concentrated on the triplet $(D, S_1, S_2)$. Chao and Lee (1992) introduced an estimator based on the estimation of both the expected sample coverage and the variation coefficient of the $q_i$'s. Chao and Lin (2012) also considered lower bounds estimators in nonparametric models, but in contrast to the aforementioned papers, they assume a sampling scheme without replacement.

Another approach is to assume that the $A_i$'s are independent variates with the same distribution $p = (p_0, p_1, \ldots, p_n)$, so that the vector $(S_0, S_1, \ldots, S_n)$ has a multinomial distribution with parameters $N$ and $p$, and the zero-truncated counts are i.i.d. with distribution given by

$$p_l^+ = \frac{p_l}{1 - p_0}, \; \text{for all integers } l \geqslant 1,$$

conditionally on the number $D$ of observed species (see Lemma 1 below). In this setting, decomposing the likelihood as a product of a term that depends only on $N$ and $p_0$, and a term that depends only on $p_j/(1 - p_0)$, $j \geqslant 1$, Sanathanan (1972) pointed out that if $p$ were known, then the maximum likelihood estimator of $N$ would be

$$\widehat{N}_p = \lfloor D/(1-p_0) \rfloor, \tag{3}$$

where $\lfloor x \rfloor$ denotes the integer part of $x$. Postulating a parametric assumption on $p$ in order to make $p_0$ identifiable, Sanathanan (1972) computed the asymptotic distribution of both the maximum likelihood estimator and the so-called conditional maximum likelihood estimator of $N$ as $N \to \infty$. The results are obtained under classical regularity assumptions on the parametric model.

Most papers adopting the point of view of independent $A_i$'s with common distribution $p$ assume, moreover, that each $A_i$ is distributed as a Poisson with expectation $\lambda_i$, the $\lambda_i$'s being independent variables from some distribution $\omega$ over $(0,\infty)$ that is called a mixing distribution. Therefore,

$$P(A_i = j) = p_j(\omega) = \int_0^\infty \frac{\lambda^j \exp(-\lambda)}{j!} d\omega(\lambda) \tag{4}$$

and such a setting is called the Poisson mixture setting. It is generally referred as parametric if a parametric assumption is formulated on $\omega$, and nonparametric otherwise.

In the parametric Poisson mixture setting, Chao and Bunge (2002) estimated $N$ by the number $\sum_{j \geqslant 2} S_j$ of duplicated species divided by an estimator for the proportion of duplications in a sample. The estimator is shown to be consistent in the case where $\omega$ is a Gamma distribution. An extension of this estimator based on the first three capture counts was proposed by Lanumteang and Böhning (2011).

Laird (1978) proved that the nonparametric maximum likelihood estimator (MLE) of a mixing distribution is typically discrete with a finite number of points of support, but no closed-form solution exists for the MLE and the number of points of support is not even known in advance. See also Lindsay (1995) for a review of the MLE properties in this context. The nonparametric Poisson mixture setting would enter the setting of Laird (1978) if the abundance were completely observed. However, this is not the case since only zero-truncated counts are available. Nevertheless, the nonparametric MLE of $(N, \omega)$ has been investigated by several authors in the nonparametric Poisson mixture setting (note that $\omega$ is identifiable provided that $\omega$ has no mass on zero, see Lemma 2.1 in Mao and Lindsay (2007)). In this setting, Norris and Pollock (1998) developed the MLE of $(N, \omega)$ on real data examples as well as on simulations. They calculated the MLE using an analogous EM-algorithm as that used by Norris and Pollock (1996) for binomial and censored geometric mixtures in the context of capture-recapture data. They proposed bootstrap-based tests (a test statistic being proposed, the critical value is evaluated by bootstrap) and estimators for classical ecological diversity and evenness measures. Böhning and Schön (2005) considered an alternative EM-algorithm for estimating iteratively $\omega$ and $N$. They selected the number of points of support for the MLE of $\omega$ by using either the AIC or the BIC criterion. Assuming that the MLE of $N$ is asymptotically Gaussian, they calculated confidence intervals for $N$ using bootstrap. Wang and Lindsay (2005) pointed out the numerical instability of earlier estimation methods and proposed to add a penalty term to the log-likelihood function in order to stabilize the estimation procedure. Their main focus in on testing for homogeneity vs. heterogeneity, that is, testing the null hypothesis of a degenerate mixing distribution $\omega$. In a more recent paper, Wang (2010) considered a continuous estimator for $\omega$, that is coined "smooth nonparametric MLE", in order to better capture the information of species abundance near zero. For example, he considered a Poisson-compound gamma model where the distribution $\omega$ is modeled by a gamma-mixture distribution parametrized by a shape parameter. He imposed an exponential prior for the odds in order to stabilize the procedure, and used an empirical Bayes method for maximizing the likelihood and a cross-validation procedure for estimating the shape parameter.

Unfortunately, there is no theoretical result on the asymptotic distribution of the aforementioned estimators in the setting of the nonparametric Poisson mixture model. In some sense, Mao and Lindsay (2007) proved that no limit distribution theory could be achievable in this setting. To be more specific, note first that as a consequence of (3), estimating $N$ amounts to estimate $p_0/(1 - p_0)$, the odds that a species is undetected in a sample. Mao and Lindsay (2007) proved the discontinuity of the odds as a function of $\omega$, from which they derived that the odds has no locally unbiased and locally informative estimator. They proved that asymptotically valid (as $D \to \infty$) confidence intervals for the odds are necessarily one-sided, which means that only lower bounds (for the odds as well as for $N$) can be calculated.

In this paper, we propose a new nonparametric approach for estimating $N$. Similar to the Poisson mixture setting, we assume that the abundances $A_i$ are independent with common distribution $p$. However, in contrast to the Poisson mixture setting, we do not assume that the common distribution is a mixture of Poisson distributions. Instead, we assume that $p$ is a convex distribution. The characterization of convex distributions as a mixture of triangular distributions allows us to assume in fact that $p$ is a *convex abundance distribution* (a term coined in Durot, et al. (2013)), which means that the first triangular component $T_1$ is absent in the mixture. This is a natural assumption since this component corresponds to a Dirac mass at zero and would therefore refer to absent species in the whole population, as the only count that could ever be observed for them is 0. Our assumption of a convex abundance distribution is rather weak but is sufficient to make identifiable the problem of abundance estimation, and to derive an estimate $\widehat{\theta}$ of $\theta = 1/(1 - p_0)$ for which the asymptotic distribution can be computed. Inspired by (3), we deduce $\widehat{N} = \lfloor D\widehat{\theta} \rfloor$ as an estimate of $N$. We provide the asymptotic distribution of $\widehat{N}$, so we are able to build confidence intervals for $N$ based on this estimator. The method is easy to implement, and the calculation of the estimators does not depend on any tuning parameter.

The shape constraint that $p$ is a convex abundance distribution is a mild assumption that does not seem to be strongly violated by data sets we have analyzed in ecology. Let us illustrate this on a few well-known examples. We will consider the Malayan butterfly data from Fisher et al. (1943), the bird abundance data considered by Norris and Pollock (1998), the tomato flower data taken from Mao and Lindsay (2003), and finally the mi-crobial species data treated by Wang (2010). In these examples, the distribution $p$ cannot be observed since the abundances are zero-truncated, but the zero-truncated distribution $p^+$ can be observed. Under our assumption of a convex abundance distribution $p$, $p^+$ is convex. Figure 1 shows that, for the considered examples, the projection of the observed zero-truncated abundance distribution on the space of convex distributions suits the data well, so that the convexity assumption is reasonable.

The paper is organized as follows. The setting is precisely defined in Section 2. Estimators for $\theta$ and $N$ and their asymptotic distributions are given in Section 3. Confidence intervals for $N$ are given in Section 4. A simulation study is reported in Section 5 to assess the performances of our estimators and confidence intervals. Finally, we compare our method to competitors on the four examples presented before in Section 6. The proofs are postponed to Section 7.

## 2 Model

### 2.1 The statistical problem

We consider a population composed of $N$ species and we assume that the data are coming from $N$ independent and identically distributed random variables $(A_1, \ldots, A_N)$, where $A_i$ is the abundance (that is, the number of individuals) of species $i$ in a sample. The distribution of $A_i$ is denoted by $p$, so that $p_j = P(A_i = j)$ for all integers $j \geqslant 0$. In fact, only species that are present in the sample can be counted, which means that species for which $A_i = 0$ are not observed. Thus, we only observe the zero-truncated counts $X_1, \ldots, X_D$, where $D$ denotes the total number of observed species in the sample. The setting can be formalized as follows:

**Lemma 1** *We observe $X_1, \ldots, X_D$, where $D$ is a binomial variable with parameters $N$ and $1 - p_0$, and conditionally on $D$, $X_1, \ldots, X_D$ are i.i.d. random variables with distribution $p^+$ defined by*
$$p_j^+ = \frac{p_j}{1 - p_0}, \text{ for all integers } j \geqslant 1. \tag{5}$$

Based on the observations $X_1, \ldots, X_D$ we aim at estimating $N$, the total number of species.

### 2.2 The assumption of a convex abundance distribution

Without any modelling assumption on the distribution $p$, $N$ is not identifiable. To make $N$ identifiable, we propose a nonparametric modelling of $p$, assuming that $p$ is a convex abundance distribution, as defined in Definition 1 below.

To motivate our Definition 1, let us first recall that in various data sets we have analysed in ecology, the assumption that $p$ is a convex discrete distribution is reasonable, see Section 6 for examples. Thus, we assume that $p$ is a convex discrete distribution on $\mathbb{N}$, which means that
$$p_i - p_{i-1} \leqslant p_{i+1} - p_i \text{ for all } i \geqslant 1.$$

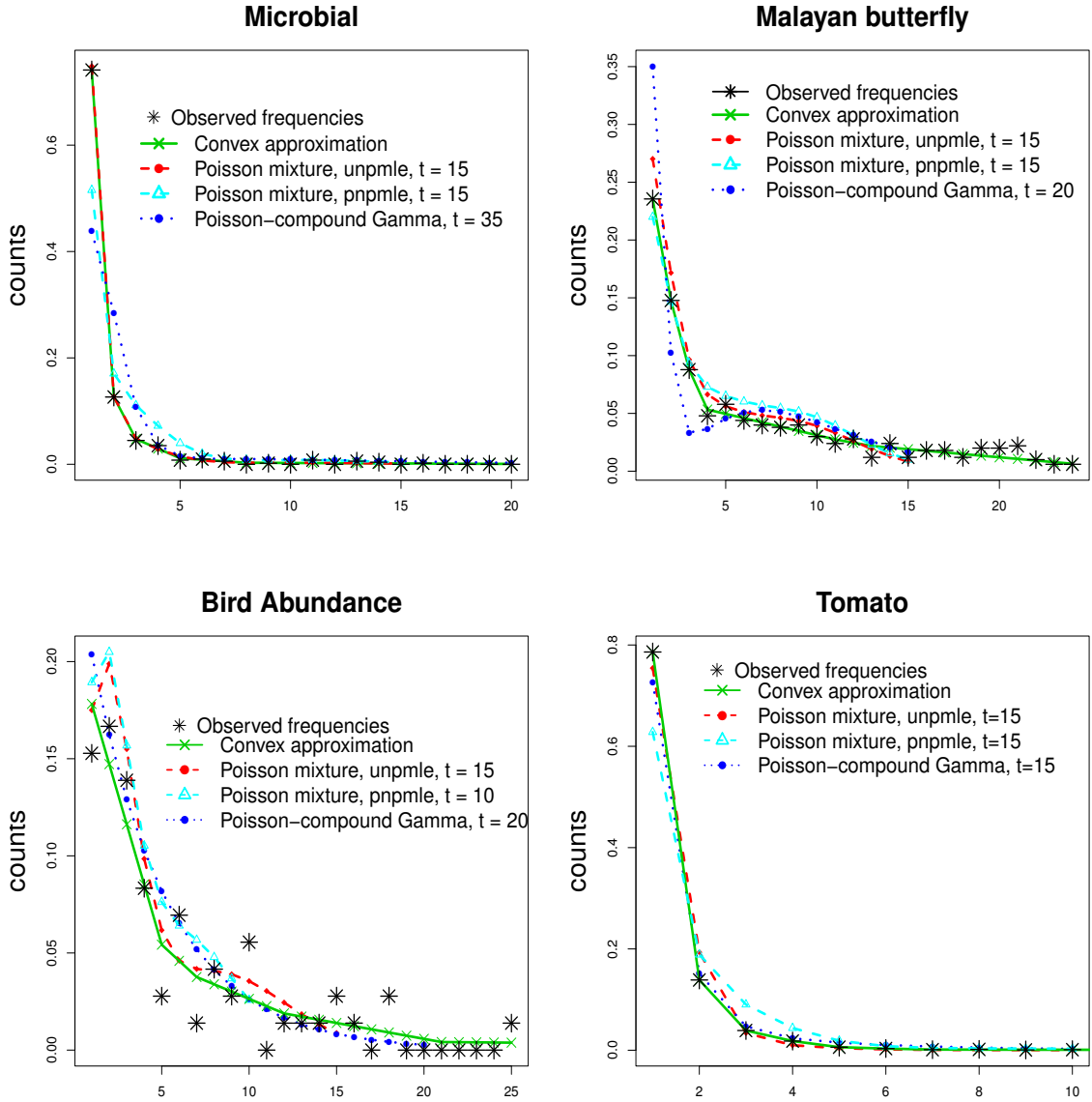

Figure 1: Estimation of the zero-truncated distribution $p^+$. The green curve is the projection of the empirical frequencies onto the set of convex distributions, the red and the light blue curves are the estimated distribution under a nonparametric Poisson mixture, and the dark blue curve is the estimation obtained under the Poisson-compound Gamma model. For the last three methods the cutoff value $t$ is given.

Note that $p$ being assumed convex on $\mathbb{N}$, $p$ is also non-increasing on $\mathbb{N}$. It follows from Theorem 7 in Durot et al. (2013) that $p$ can be decomposed into a mixture of triangular distributions, and that this mixture is unique. More precisely,

$$p_i = \sum_{j \geqslant 1} \pi_j T_j(i), \tag{6}$$

for all integers $i \geqslant 0$, where

$$\pi_j = \frac{j(j+1)}{2}(p_{j+1} + p_{j-1} - 2p_j) \text{ for all integers } j \geqslant 1 \tag{7}$$

and where $T_j$ is the triangular distribution defined by

$$T_j(i) = \begin{cases} \frac{2(j-i)}{j(j+1)} & \text{for all } i \in \{0, \ldots, j-1\} \\ 0 & \text{for all integer } i \geqslant j. \end{cases}$$

Our interpretation of the mixture (6) is that the set of species is separated into groups, each species having probability $\pi_j$ to belong to the group $j$ of species, and the abundance distribution of all species in the group $j$ is the triangular distribution $T_j$. As the first component $T_1$ is a Dirac mass in 0, it refers to species for which the only abundance that could be observed is 0. This group simply defines absent species, and therefore $\pi_1$ has to be zero in an abundance distribution. This leads us to the following definition:

**Definition 1** *The distribution $p$ on $\mathbb{N}$ is a convex abundance distribution if there exist positive weights $\pi_j$, $j \geqslant 2$ such that $p_i = \sum_{j \geqslant 2} \pi_j T_j(i)$ for all integers $i \geqslant 0$.*

In the sequel, we assume that the abundance distribution $p$ is a convex abundance distribution. It then follows from Equation (7) that

$$p_2 + p_0 - 2p_1 = 0, \tag{8}$$

or equivalently,

$$\frac{1}{1-p_0} = 2p_1^+ - p_2^+ + 1, \tag{9}$$

where $p^+$ is the zero-truncated distribution defined by (5). The distribution $p^+$ is identifiable since we observe $X_1, \ldots, X_D$ which are i.i.d. with distribution $p^+$ conditionally on $D$. Therefore, it follows from (9) that $1 - p_0$ is identifiable and because $D$ has a binomial distribution with parameters $N$ and $1 - p_0$, we conclude that $N$ also is identifiable. This proves that our assumption is sufficient to avoid identifiability problems. The precise construction of the estimates is the aim of the following section.

# 3 Estimators for $\theta$ and $N$

In order to estimate $N$, we first build an estimator for

$$\theta = \frac{1}{1 - p_0}. \tag{10}$$

Because of (9), we consider estimators of the form

$$\widehat{\theta} = 2\widehat{p}_1^+ - \widehat{p}_2^+ + 1, \tag{11}$$

where $\widehat{p}^+$ is a given estimator for $p^+$. Then, inspired by (3), we estimate $N$ by

$$\widehat{N} = D\widehat{\theta}. \tag{12}$$

In this section, we study two different estimators for $p^+$ which result in two different estimators for $\theta$ and $N$. We provide the asymptotic distribution of both considered estimators for $N$.

## 3.1 Estimators based on the empirical estimator of $p^+$

The more commonly used estimator for a discrete distribution is the empirical estimator. In our case, the empirical estimator $f$ of $p^+$ is defined by

$$f_j = \frac{1}{D} \sum_{i=1}^{D} I(X_i = j) = \frac{S_j}{D} \tag{13}$$

for all integers $j \geqslant 1$, where we recall that $S_j$ denotes the number of species having abundance $j$, see (1). Using this estimator in (11) and (12) leads to the estimators $\widehat{\theta}_f = 2f_1 - f_2 + 1$ and

$$\widehat{N}^f = D\widehat{\theta}_f = 2S_1 - S_2 + D. \tag{14}$$

The asymptotic distribution of this estimator is easy to compute: one can derive from the central limit theorem that

$$\frac{\widehat{N}^f - N}{\sqrt{6S_1}} \text{ converges in law to } \mathcal{N}(0, 1), \tag{15}$$

see Section 7 for details.

## 3.2 Estimators based on the constrained least-squares estimator of $p^+$

The estimator (14) exploits the convexity assumption only through the identity (9). On the other hand, it immediatly follows from its definition that $p^+$ is convex under our assumption. We might obtain better estimates by incorporating this information into our estimation procedure, so instead of the empirical estimator, we consider here a convex estimator of $p^+$. Precisely, we consider the constrained least-squares estimator $\widehat{p}^+$ of $p^+$, defined as the unique solution to the following optimisation problem:

$$Q(\widehat{p}^+) = \inf_{q \in \mathcal{C}} Q(q), \quad \text{where } Q(q) = \sum_{j \geqslant 1}(q_j - f_j)^2 \qquad (16)$$

and where $\mathcal{C}$ denotes the set of all convex sequences $q$ on $\mathbb{N}$ having $\sum_{j \geqslant 1} q_j^2 < \infty$.

It follows from the results in Durot et al. (2013) that $\widehat{p}^+$ exists, has a finite support, and is a probability mass function. Thus, $\widehat{p}^+$ is the convex probability function that is the closest to $f$ in the $\ell_2$-sense. Moreover, an algorithm for computing $\widehat{p}^+$ in a finite number of steps is described in Durot et al. (2013). It is based on the support reduction algorithm proposed by Groeneboom et al. (2008).

In the sequel, we denote by $\widehat{\theta}$ and $\widehat{N}$ the estimators defined by (11) and (12), where $\widehat{p}^+$ denotes the constrained least-squares estimator defined by (16).

## 3.3 Asymptotic distributions of $\widehat{\theta}$ and $\widehat{N}$

To compute the limit distributions of $\widehat{\theta}$ and $\widehat{N}$, we need to introduce the following definition.

**Definition 2** *(i). An integer $i \geqslant 2$ is called a knot of $p^+$ if $p_i^+ - p_{i-1}^+ < p_{i+1}^+ - p_i^+$.*

*(ii). An integer $i \geqslant 2$ is called a double-knot of $p^+$ if both $i$ and $i+1$ are knots of $p^+$.*

Let us notice that $p^+$ necessarily has at least one finite knot since it is a convex probability mass function. However, double-knots of $p^+$ may not exist.

Let us introduce some more notation.

- Let $\tau$ be the maximum of the support of $p^+$ if $p^+$ has a finite support, and $\tau = \infty$ otherwise.

- Let $\kappa$ be the smallest double-knot of $p^+$ if $p^+$ has at least one double-knot, and let $\kappa = \infty$ otherwise.

- For a given integer $k \geqslant 2$ and a given set $\mathcal{I} \subset \{1, \ldots, k\}$ that contains 1 and $k$, let $\mathcal{C}^{\mathcal{I}}$ be the set of sequences $q \in \mathbb{R}^k$ that are convex in all $i \notin \mathcal{I}$, with no constraint at points $i \in \mathcal{I}$, and let $\Phi_{\mathcal{I}}$ be the function defined for all vectors $t = (t_1, t_2, \ldots, t_k) \in \mathbb{R}^k$ by

$$\Phi_{\mathcal{I}}(t) = \arg\min_{q \in \mathcal{C}^{\mathcal{I}}} \sum_{j=1}^{k}(q_j - t_j)^2.$$

To be more formal, denoting by $1 = i_1 < i_2 < \ldots < i_I = k$ the points in $\mathcal{I}$, $\mathcal{C}^{\mathcal{I}}$ is defined by

$$\mathcal{C}^{\mathcal{I}} = \left\{ q \in \mathbb{R}^k \text{ such that } q \text{ is convex on } \{i_{j-1}, \ldots, i_j\} \text{ for all } j = 2, \ldots, I \right\}.$$

Note that $\mathcal{C}^{\mathcal{I}}$ is a closed convex cone in $\mathbb{R}^k$, so $\Phi_{\mathcal{I}}(t)$ is uniquely defined for all $t \in \mathbb{R}^k$.

In what follows, we assume that either $p^+$ has a finite support, or $p^+$ has at least one double knot. This amounts to assume that $\min\{\tau, \kappa\} < \infty$. The following theorem proves that both limit distributions of $\widehat{\theta}$ and $\widehat{N}$ depend on the distribution of $\Phi_{\mathcal{I}}(W)$ for a given set $\mathcal{I}$ and a given Gaussian vector $W$.

**Theorem 1** *Let $\widehat{\theta}$ and $\widehat{N}$ be defined by (11) and (12), with $\widehat{p}^+$ defined by (16). Assume that $p^+$ is convex with*

$$\min\{\tau, \kappa\} < \infty \qquad (17)$$

*and consider a finite integer $k \geqslant \min\{\tau + 1, \kappa\}$. Denoting by $W$ a centered Gaussian vector in $\mathbb{R}^k$ with covariance matrix $\Gamma$ defined as $\Gamma_{jj} = p_j^+(1 - p_j^+)$ and $\Gamma_{jj'} = -p_j^+ p_{j'}^+$ for all $1 \leqslant j, j' \leqslant k$ and $j < j'$, and by $\mathcal{I} = \{1, k\} \cup \mathcal{J}$ where $\mathcal{J}$ is the set of knots of $p^+$ that are smaller than $k$, we have:*

(i). $\sqrt{D}\left(\widehat{\theta} - \theta\right)$ *converges in law to* $2\Phi_{\mathcal{I}}(W)_1 - \Phi_{\mathcal{I}}(W)_2$ *as $N$ goes to infinity,*

(ii). $(\widehat{N} - N)/\sqrt{D}$ *converges in law to* $2\Phi_{\mathcal{I}}(W)_1 - \Phi_{\mathcal{I}}(W)_2 + T$ *as $N$ goes to infinity, where $T$ is a $N(0, \theta(\theta - 1))$ variable independent of $\Phi_{\mathcal{I}}(W)$.*

Let us give a few comments:

- The fact that $k$ can be chosen in an arbitrary way provided that $k \geqslant \min\{\tau + 1, \kappa\}$ could be of practical interest. Indeed, if one wants to estimate the asymptotic distribution of either $\widehat{\theta}$ or $\widehat{N}$, then one has to choose a convenient value for $k$ and under our assumptions, any large enough $k$ suits (the precise value of $\min\{\tau + 1, \kappa\}$ need not to be known, and only an upper bound is needed). Let us notice that the choice $k = \widehat{\tau}^f + 1$, where $\widehat{\tau}^f$ is the maximum of the support of the empirical estimator, suits in all cases where the support of $p^+$ is finite. In order to save computational time, one can choose a smaller $k$ provided that one is confident that the choosen $k$ has $k \geqslant \kappa$.

- We have assumed for simplicity that $\kappa$ is the smallest double-knot of $p^+$ (if $p^+$ has some) but $\kappa$ could in fact denotes any double-knot of $p^+$ (if $p^+$ has some).

- If $p^+$ is a triangular distribution, then $\tau$ is finite, $k \geqslant \tau + 1$, and we have $p_j^+ = 0$ for all $j \geqslant \tau + 1$ so that the $j$-th component of $W$ is almost surely equal to 0 for all $j = \tau + 1, \ldots, k$.

The limit distribution of the convex least squares estimator of a discrete distribution has been established by Balabdaoui et al. (2014) in the case where the true distribution has a finite support. Their results could be used to prove our Theorem 1 in that case, but this would not be straightforward since in our case, the empirical estimator of $p^+$ defined by Equation (13) is based on a random number $D$ of observations. Moreover, this would not cover the case of an infinite support. Therefore, we provide a complete and original proof of Theorem 1 in Section 7.3.

It is worth mentioning that whereas the limit distributions of $\widehat{\theta}$ and $\widehat{N}$ take a quite complicated form in the general case, those limit distributions are Gaussian in the particular case where $\min\{\tau+1,\kappa\}=2$, that is, when either $p^+$ is the dirac measure at point 1, or $p^+$ has a double knot at point 2. To be more precise, note that considering $k=2$ yields $\Phi_{\mathcal{I}}(W)=W$, so that $2\Phi_{\mathcal{I}}(W)_1 - \Phi_{\mathcal{I}}(W)_2 = 2W_1 - W_2$. Thus, $2\Phi_{\mathcal{I}}(W)_1 - \Phi_{\mathcal{I}}(W)_2$ as well as $2\Phi_{\mathcal{I}}(W)_1 - \Phi_{\mathcal{I}}(W)_2 + T$ are Gaussian variables. Thanks to Equality (9) the variance of these variables can be computed according to the following corollary.

**Corollary 1** *In the particular case where $\min\{\tau+1,\kappa\}=2$ we have:*

(i). $\sqrt{D}\left(\widehat{\theta} - \theta\right)$ *converges in law to* $\mathcal{N}(0, 6p_1^+ - \theta(\theta-1))$ *as $N$ goes to infinity,*

(ii). $(\widehat{N} - N)/\sqrt{D}$ *converges in law to* $\mathcal{N}(0, 6p_1^+)$ *as $N$ goes to infinity.*

If we knew in advance that $\min\{\tau+1,\kappa\}=2$, then the limit distributions of $\widehat{\theta}$ and $\widehat{N}$ would be easy to estimate, using the estimator $\widehat{p}^+$ for $p^+$ together with the estimator $\widehat{\theta}$ for $\theta$. Unfortunately, $\min\{\tau+1,\kappa\}$ is not known in advance so one has to estimate the limit distributions in the general case in order to build confidence intervals. This is developed in the following section.

## 4 Confidence intervals

In this section, we investigate several constructions of confidence intervals for $N$.

### 4.1 Estimation based on the empirical frequencies

If $N$ is large and if the quantities $Np_1$ and $N(1-p_1)$ are not too small, then it follows from (15) that a confidence interval for $N$ can be calculated assuming that the distribution of $(\widehat{N}^f - N)/\sqrt{6S_1}$ can be approximated by that of a standard Gaussian variable. This leads to the following confidence interval

$$\mathrm{CI}^f = \left[\widehat{N}^f - \nu_{1-\alpha/2}\sqrt{6S_1} \;,\; \widehat{N}^f + \nu_{1-\alpha/2}\sqrt{6S_1}\right], \tag{18}$$

where $\alpha \in (0,1)$ is fixed and $\nu_{1-\alpha/2}$ is the $(1-\alpha/2)$-quantile of the standard Gaussian law. According to (15), the asymptotic level of the interval is $1-\alpha$.

### 4.2 The plug-in procedure on $\widehat{N}$

The limit distribution of $\widehat{N}$ given in Theorem 1 depends on $p^+$ through $k$, $\mathcal{I}$, the covariance matrix $\Gamma$ of the Gaussian vector $W$, and the variance $\theta(\theta-1)$ of $T$. We estimate $\theta(\theta-1)$ by $\widehat{\theta}(\widehat{\theta}-1)$ where $\widehat{\theta}$ is defined in (11) with $\widehat{p}^+$ the constrained least-squares estimator of $p^+$, and we estimate all unknown quantities depending on $p^+$ by similar quantities with $p^+$ replaced by $\widehat{p}^+$. For simplicity, we consider $k = \min\{\tau+1,\kappa\}$ and we estimate $k$ and $\mathcal{I}$ as follows. Let us denote by $\widehat{s}$ the first double knot of $\widehat{p}$ if it exists. In the case where such a double knot does not exist we use the convention that $\widehat{s} = \infty$. Let $\widehat{\tau}$ be the maximum of the support of $\widehat{p}^+$. From Theorem 1 in Durot et al. (2013) we know that $\widehat{\tau}$ is finite. Therefore, $\widehat{k} = \min\{\widehat{s}, \widehat{\tau} + 1\}$ is finite and $\mathcal{I}$ is estimated by $\widehat{\mathcal{I}}$, the set consisting of 1 and the knots of $\widehat{p}^+$ before $\widehat{k}$. Then, the estimated quantiles of the random variable $2\Phi_\mathcal{I}(W)_1 - \Phi_\mathcal{I}(W)_2 + T$, say $\widehat{\lambda}_{1-\alpha/2}$ and $\widehat{\lambda}_{\alpha/2}$, are calculated by simulation. The calculation of $\Phi^\mathcal{I}(W)$ is done using the algorithm proposed by Dykstra (1983) for restricted least squares regression; see Balabdaoui et al. (2014) for more details. Then we consider the confidence interval for $N$ given by

$$\text{CI} = \left[\widehat{N} - \widehat{\lambda}_{1-\alpha/2}\sqrt{D}, \widehat{N} - \widehat{\lambda}_{\alpha/2}\sqrt{D}\right]. \tag{19}$$

The obvious advantage of the interval (18) as compared to (19) lies in its computational simplicity, due to the fact that it is based on the empirical frequencies rather than on the constrained estimator $\widehat{p}^+$. However, it is not clear in the general case which of those two intervals has better length or coverage probability. Such a comparison can easily be performed only in the particular case where $\min\{\tau + 1, \kappa\} = 2$. Indeed, since $D$ is distributed as a Binomial variable with parameters $N$ and $1 - p_0 = 1/\theta$, it follows from Corollary 1 that $(\widehat{N} - N)/\sqrt{N}$ converges in law to $\mathcal{N}(0, 6p_1)$ as $N$ goes to infinity. This can be compared to the similar convergence result (22) for $\widehat{N}^f$: in the case where $k = 2$, $\widehat{N}$ and $\widehat{N}^f$ have the same limit distribution, so the difference between the two intervals mainly relies in the way we estimate the unknown parameters in the limit distribution, and on the chosen center $\widehat{N}_f$ or $\widehat{N}$ for the interval . The comparison between these two intervals will be studied in the next section.

An alternative to the plug-in method is to use a bootstrap procedure for estimating the quantiles of the limit distribution of $\widehat{N}$.

### 4.3 The bootstrap procedure on $\widehat{N}$

The bootstrap procedure consists in creating a bootstrap sample as follows: we first draw $D^*$ as a binomial variable with parameters $\widehat{N}$ and $1/\widehat{\theta}$. Then we draw $(X_1^*, \ldots, X_{D^*}^*)$, a $D^*$-sample with distribution $\widehat{p}^+$.

We calculate the statistics $f_j^* = \sum_{i=1}^{D^*} I(X_i^* = j)/D^*$ and the bootstrap estimator of $p^+$ by minimizing $\sum_{j \geqslant 1}(q_j - f_j^*)^2$ over $q \in \mathcal{C}$. Finally we get $\widehat{\theta}^*$, the bootstrap estimator of $\theta$. For a fixed $\beta \in (0,1)$, the $\beta$-quantile $\zeta_\beta$ of $(\widehat{N} - N)/\sqrt{D}$ is estimated by the $\beta$-quantile $\zeta_\beta^*$ of the distribution of $(D^*\widehat{\theta}^* - \widehat{N})/\sqrt{D^*}$. Finally the bootstrap confidence interval for $N$ is written

$$\text{CI}^* = \left[\widehat{N} - \sqrt{D}\zeta_{1-\alpha/2}^*\,,\,\widehat{N} - \sqrt{D}\zeta_{\alpha/2}^*\right], \tag{20}$$

## 5 Simulation study

We designed a simulation study to assess the performances of the proposed estimators and of the associate confidence intervals for $N$. We considered the convex estimator $\widehat{N}$ and the empirical one $\widehat{N}^f$, and the three confidence intervals $\text{CI}^f$, CI and $\text{CI}^*$ defined respectively at Equations (18), (19), (20). In the second part of this study we compare our procedure to other methods already proposed in the literature.

## 5.1 Simulation design

**Convex abundance distribution** We considered a Poisson mixture setting, which means that the distribution $p$ in our simulations takes the form (4), where we considered a Gamma mixing distribution $\omega$. For this choice, we were motivated by the fact that all methods to which we will compare are either based on that assumption, or are proved to give consistent estimators of $N$ under this assumption, or consider a statistical modeling that covers this distribution. Precisely, $p$ is a Gamma-Poisson distribution that takes the form

$$p_j = \frac{\Gamma(j+\nu)}{\Gamma(\nu)j!}\mu^\nu(1-\mu)^j \qquad (21)$$

for some unknown $\nu > 0$ and $\mu \in (0,1)$. Note that for such distributions, $p_0 = \mu^\nu$ and $p_j^+ = p_j/(1-\mu^\nu)$ for $j \geqslant 1$.

We focused on the case where $p$ is a convex abundance distribution which is satisfied when

$$\nu > 1 \quad \text{and} \quad 1-\mu = \frac{2\nu - \sqrt{2\nu(\nu-1)}}{\nu(\nu+1)},$$

(see Section 7.4 for a proof of this result).

**Simulation parameters and evaluation criteria** To cover a wide range of possible applications to ecological, microbial and other similar data, we choose $N$ in the set

$$\{50, 100, 200, 400, 800, 1500, 3000, 5000, 10000\}.$$

We considered several values of $\nu$, $\nu \in \{1.01, 1.05, 1.1, 1.3, 1.5, 1.75\}$ corresponding to the following values of $p_0$: $\{0.073, 0.16, 0.218, 0.33, 0.382, 0.42\}$. All confidence intervals are computed at level $\alpha = 0.05$.

All simulation results are based on 1000 samples. The quantiles of the bootstrap distribution – denoted $(\zeta^*_{\alpha/2}, \zeta^*_{1-\alpha/2})$ – and those of the asymptotic distribution where the unknown parameters have been replaced by their estimators – denoted $(\widehat{\lambda}_{\alpha/2}, \widehat{\lambda}_{1-\alpha/2})$ – are also calculated on the basis of 1000 simulations. The simulation were carried out with R (www.r-project.org). The R functions are available at http://w3.jouy.inra.fr/unites/miaj/public/perso/SylvieHuet_en.html.

The accuracy of an estimate $\widehat{N}$ is measured in terms of bias, standard-error and prediction error. The bias of an estimator $\widehat{N}$, defined as $N - E(\widehat{N})$, is estimated by `bias` $= N - \widehat{N}_\bullet$, where $\widehat{N}_\bullet = \sum_s \widehat{N}_s / 1000$ with $\widehat{N}_s$ being the estimate of $N$ at simulation $s$. The standard-error of $\widehat{N}$ is estimated by `se` $= \sqrt{\sum_s (\widehat{N}_s - \widehat{N}_\bullet)^2 / 1000}$, and the (squared) error of prediction by $\texttt{EP}^2 = \texttt{bias}^2 + \texttt{se}^2$.

The quality of the confidence intervals is measured in terms of non-coverage probability at each of their endpoints. Namely, for a given interval $[B_{\text{inf}}, B_{\text{sup}}]$, we estimated the left and the right non-coverage probabilities – defined respectively by $P(N < \mathrm{B}_{\inf})$ and $P(N > \mathrm{B}_{\sup})$ – by $\sum_s I(N < \mathrm{B}_{s,\inf})/1000$ and $\sum_s I(N > \mathrm{B}_{s,\sup})/1000$, respectively, where $\mathrm{B}_{s,\inf}$ and $\mathrm{B}_{s,\sup}$ stand for the bounds obtained at simulation $s$.

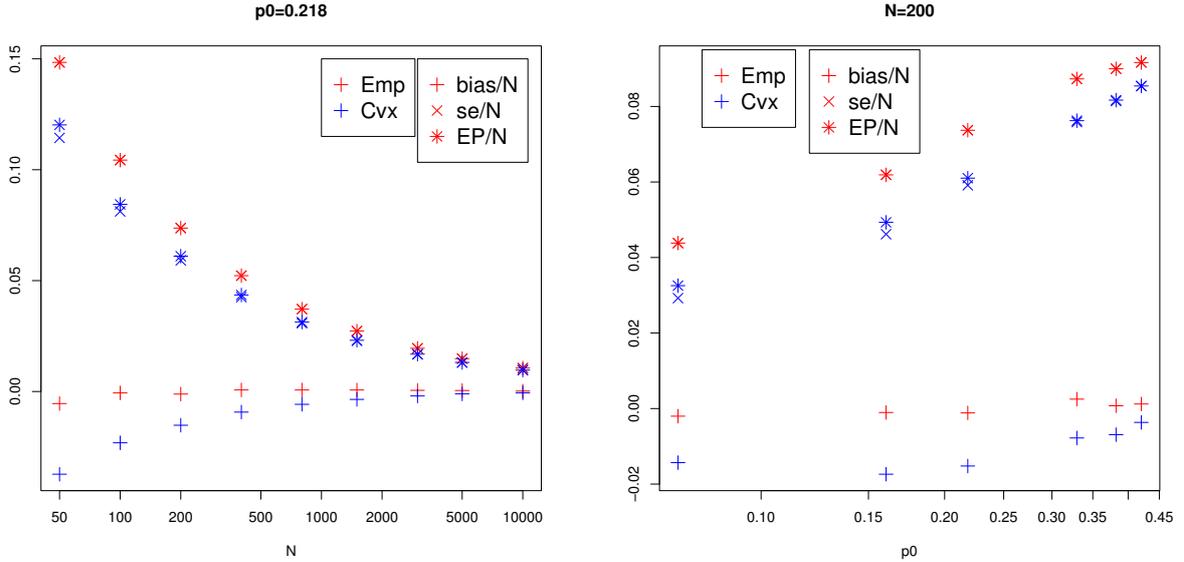

Figure 2: Relative bias, standard-errors and prediction error versus $N$ for $p_0 = 0.218$ on the left side, versus $p_0$ for $N = 200$ on the right side.

## 5.2 Comparison of $\widehat{N}$ and $\widehat{N}^f$

The relative bias, standard-error and mean squared error of prediction for both estimators $\widehat{N}$ and $\widehat{N}^f$ are shown on Figure 2. For $p_0 = 0.218$ (respectively $N = 200$), we display the relative quantities `bias`/$N$, `se`/$N$ and `EP`/$N$ versus $N$ (respectively versus $p_0$). The graphs for other values of $p_0$ and $N$, being similar to those two, are omitted.

While $\widehat{N}^f$ is nearly unbiased, $\widehat{N}$ tends to over estimate $N$ for small values of $N$ and $p_0$. However, $\widehat{N}$ has a smaller standard-error than $\widehat{N}^f$, and finally a smaller prediction error. Both estimators become more accurate when $N$ increases, and when $p_0$ decreases. Indeed, when $p_0$ is small, almost all the species have been observed, leading to a smaller value of the standard-error.

## 5.3 Comparison of confidence intervals

The estimated non-coverage probabilities of the confidence intervals are given at Figures 3 and 4. We remind that they are to be compared with 2.5%. Note however that the standard-error of the estimated non-coverage probabilities based on 1000 simulations equals

$$\sqrt{\frac{0.025(1-0.025)}{1000}} \approx 0.5\%,$$

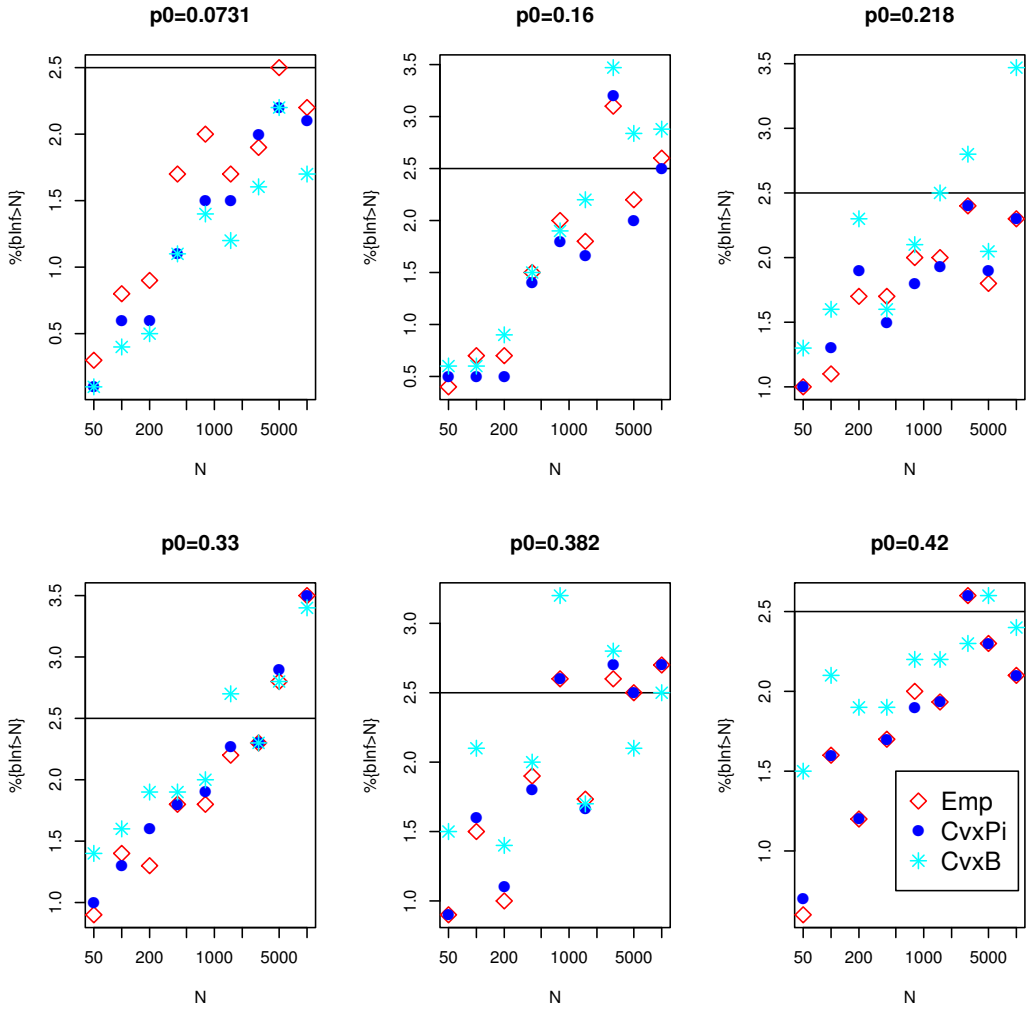

Figure 3: Comparison of confidence intervals. Estimated values of $P(N < B_{\text{inf}})$ versus $N$ for each value of $p_0$. The legend is the following : $\diamond$ is for $\text{CI}^f$ (see Equation (18)), $\bullet$ is for CI (see at Equation (19)), and $*$ is for the bootstrap confidence interval $\text{CI}^*$ (see Equation (20)).

so the estimated non-coverage probabilities are expected to lie typically within 1.5% and 3.5%.

- *Lower bound of the confidence intervals.* For the smallest values of $p_0$, the estimated values of $P(N < B_{\text{inf}})$ are smaller than 2.5%, which means that the lower bounds of the confidence intervals are too conservative. This tendancy vanishes when $N$ increases. The three methods are nearly equivalent.

- *Upper bound of the confidence intervals.* For the largest values of $p_0$, the estimated values of $P(N > B_{\text{sup}})$ are greater than 2.5%, and tend to decrease with $N$ up to 2.5%. This means that when $N$ is small, the upper bounds of the confidence intervals are too small. For the smallest values of $p_0$, the interval $\text{CI}^f$ based on the empirical procedure gives very high values of the non-coverage probability when $N$ is small. When $N$ increases, the estimated $P(N > B_{\text{sup}})$ fluctuate around 4%. To verify if these values tend to 2.5% when $N$ increases, we completed the simulation study by considering $N = 10^5$, respectively $N = 5 \times 10^5$, with $p_0 = 0.218$. The estimated values of of the non-coverage probabilities $P(N > B_{\text{sup}})$ are then equal to 2.6%, respectively 2.4%, for the interval CI based on the plug-in procedure.

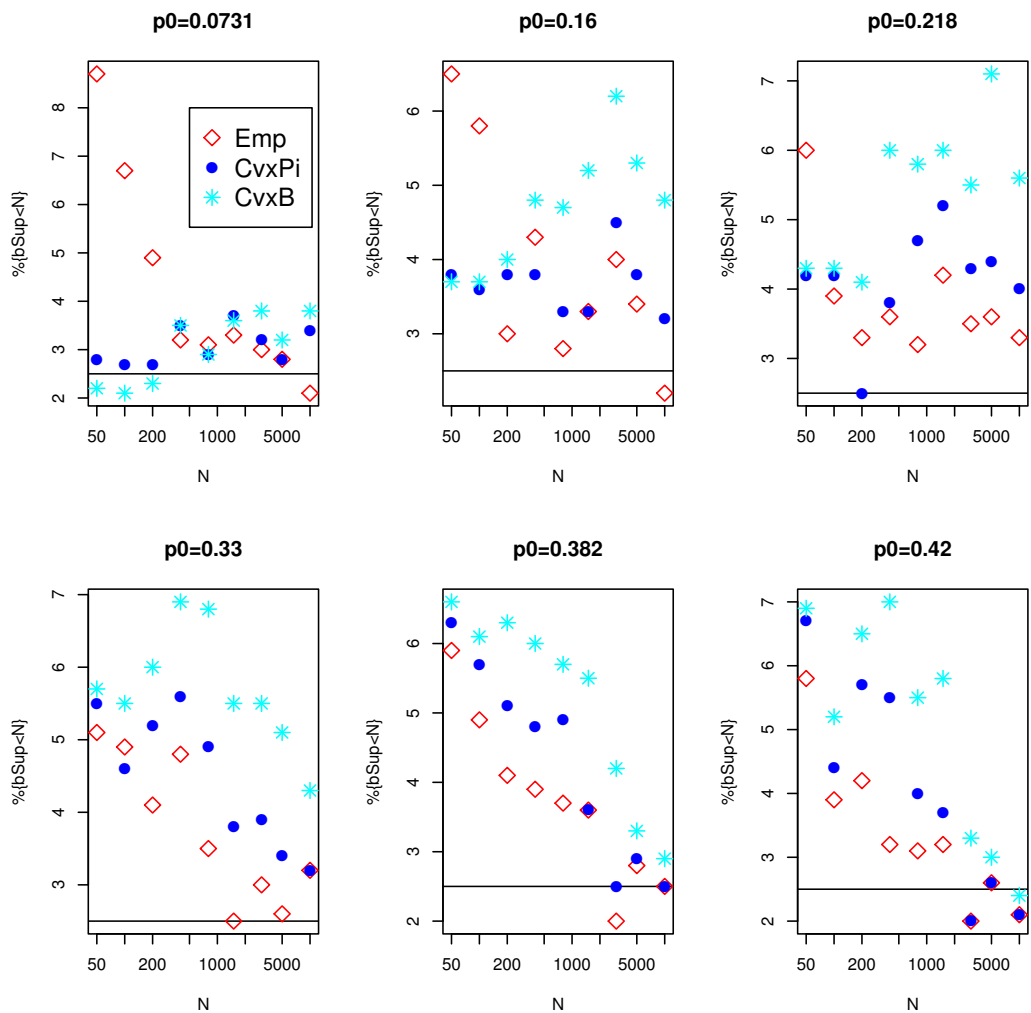

Figure 4: Comparison of confidence intervals. Estimated values of $P(N > B_{\text{sup}})$ versus $N$ for each value of $p_0$. The legend is the following : $\diamond$ is for $\text{CI}^f$ (see Equation (18)), $\bullet$ is for CI (see at Equation (19)), and $*$ is for the bootstrap confidence interval $\text{CI}^*$ (see Equation (20)).

Which method to choose among the empirical procedure, the bootstrap or the plug-in procedure? The interval based on the empirical procedure is very easy to calculate and gives reasonable results provided that $N$ is high and $p_0$ not too small. In some cases, the plug-in procedure gives better results, especially when $p_0$ is small. The bootstrap procedure does not seem to give better results than the plug-in in this simulation study. Therefore we recommend to use either the empirical procedure or the plug-in procedure. Obviously the computation time for the plug-in method is higher than for the empirical one, and depends on both $N$ and $\nu$. In our simulation study, the worse case is for $N = 800$ and $\nu = 1.01$ (which corresponds to $p_0 = 0.0731$), where the mean computation time over 1000 simulations is 70 s using an algorithm written in `R` on a 64 bits processor with 48 Go of RAM.

## 5.4 Comparison with other methods

Let us now compare our methods to those proposed in the litterature. In the sequel, we will denote by `Emp` the method that consists in estimating $N$ using the estimator $N^f$ and the confidence interval $\text{CI}^f$ defined respectively in (14) and (18), and we will denote by `CvxPi` the method that consists in estimating $N$ using the estimator $\widehat{N}$ based on the convex least-squares estimator, and the confidence interval CI defined in (19). Because we consider a non-parametric point of view, we focus our comparison on methods that do not need to estimate a parametric distribution of the abundance distribution. This includes the methods proposed by Chao (1984), Chao and Lee (1992), Chao and Bunge (2002), and Lanunteang and Böhning (2011) that will be denoted `chao84`, `ChaoLee`, `ChaoBunge`, `LB`. We will also consider the following methods based on the maximum likelihood estimation of $N$ and $p^+$ under the assumption of a Poisson mixture model, or a Poisson-compound Gamma model: `unpmle` proposed by Norris and Pollock (1996, 1998), `pnpmle` proposed by Wang and Lindsay (2005, 2008) and `pcg` proposed by Wang (2010). The simulation were carried out using the library `SPECIES` in `R`, Wang (2011).

As for the simulation design, we restricted our simulation study to two values of $N$, namely $N = 100$ and $N = 5000$, and two values of $\alpha$, namely $\alpha = 1.01$ and $\alpha = 1.75$. The methods available in the `R` function `pcg`, `pnpmle`, and `unpmle` failed to converge on several simulations and are therefore omitted. We will come back to the comparison with these methods in Section 6.

The results are given in Table 1. The methods `Emp` and `CvxPi` outperform the other methods in almost all considered situations in terms of the error of prediction. Moreover, the non-coverage probabilities are not too far from 2.5%, especially for $N = 5000$. They are typically much closer to 2.5% than the other methods. For the other methods, the behavior of the bias and standard-error depend strongly on the considered case. When $N$ is large, for all these methods except `ChaoBunge` with $p_0 = 0.42$, the confidence intervals are shifted to the left, the upper bound being smaller than $N$. This behavior is less marked for the `LB` method.

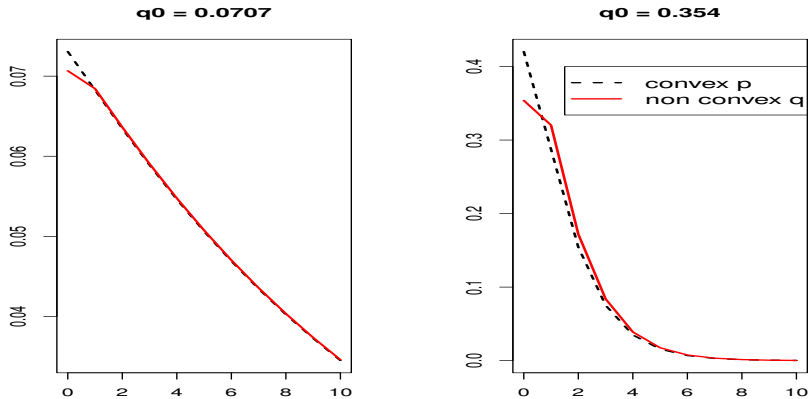

Figure 5: Representation of distributions $p$ and $q$, showing how $q$ deviates from convexity.

## 5.5 Robustness to convexity

As already noticed in the introduction, the assumption of convexity on $p^+$ seems reasonable when looking at the observed zero-truncated abundance distributions in several examples. Nevertheless, convexity of $p^+$ does not imply the convexity of $p$. To evaluate the robustness of our procedure to convexity of $p$, we carried out a simulation study considering distributions $q$ defined as follows:

$$q_0 = (p_0 + p_1)/2, \text{ and } q_j = p_j \frac{1 - q_0}{1 - p_0} \text{ for } j \geqslant 1$$

where the probabilities $p_j$, for $j \geqslant 0$ are defined at Equation (21). These distribution are represented on Figure 5.

The results are given at Table 2. As expected, our procedures lead to negatively biased estimators, particularly in the case where $q_0 = 0.354$, but the standard-errors are not affected by the lack of convexity. The confidence intervals are shifted to the right, the lower bound of the confidence interval being always larger than $N$ in the case $N = 5000$ and $q_0 = 0.354$. The behavior of the other methods depends strongly on the values of $(N, q_0)$.

## 6 Illustration on public datasets

In this section we come back to the examples presented in the introduction. For each data set we estimated the zero-truncated abundance distribution $p^+$, the number of species $N$, and a confidence interval for $N$ using all methods described in Subsection 5.4. The results are given in Table 3 and Figure 1.

Table 1: Comparison of several methods for estimating $N$: bias, standard-error, prediction error and non-coverage probabilities are reported, for $N = 100$ and $N = 5000$, considering $p_0 = 0.073$ and $p_0 = 0.42$.

|  | Emp | CvxPi | ChaoBunge | chao84 | ChaoLee | LB |
|---|---|---|---|---|---|---|
| | | | $N = 100$, $p_0 = 0.073$ | | | |
| bias/N | -0.002 | -0.02 | 0.041 | 0.02 | 0.042 | -0.31 |
| se/N | 0.061 | 0.041 | 0.033 | 0.057 | 0.031 | 2.24 |
| EP/N | 0.061 | 0.046 | 0.052 | 0.061 | 0.052 | 2.26 |
| $P(B_{\text{inf}} > N)$ | 0.8 | 0.6 | 0 | 1 | 0 | 0 |
| $P(B_{\text{sup}} < N)$ | 6.7 | 2.7 | 38 | 6.9 | 40 | 22 |
| | | | $N = 100$, $p_0 = 0.42$ | | | |
| bias/N | -0.0007 | -0.011 | -0.123 | 0.124 | 0.06 | -0.22 |
| se/N | 0.133 | 0.121 | 1.69 | 0.158 | 0.194 | 1.20 |
| EP/N | 0.133 | 0.122 | 1.69 | 0.201 | 0.203 | 1.22 |
| $P(B_{\text{inf}} > N)$ | 1.6 | 1.6 | 0 | 0.4 | 1 | 0 |
| $P(B_{\text{sup}} < N)$ | 3.9 | 4.4 | 6.5 | 13.6 | 8 | 20 |
| | | | $N = 5000$, $p_0 = 0.073$ | | | |
| bias/N | -0.0001 | -0.0025 | 0.045 | 0.036 | 0.045 | 0.016 |
| se/N | 0.0091 | 0.0066 | 0.0044 | 0.0060 | 0.0045 | 0.014 |
| EP/N | 0.0091 | 0.007 | 0.045 | 0.036 | 0.045 | 0.022 |
| $P(B_{\text{inf}} > N)$ | 2.5 | 2.2 | 0 | 0 | 0 | 0 |
| $P(B_{\text{sup}} < N)$ | 2.8 | 2.8 | 100 | 100 | 100 | 33 |
| | | | $N = 5000$, $p_0 = 0.42$ | | | |
| bias/N | 0.0002 | 0.0002 | 0.008 | 0.153 | 0.085 | 0.053 |
| se/N | 0.018 | 0.018 | 0.04 | 0.019 | 0.025 | 0.054 |
| EP/N | 0.018 | 0.018 | 0.041 | 0.154 | 0.089 | 0.076 |
| $P(B_{\text{inf}} > N)$ | 2.3 | 2.3 | 0.9 | 0 | 0 | 0.2 |
| $P(B_{\text{sup}} < N)$ | 2.6 | 2.6 | 4.3 | 100 | 89 | 23 |

Table 2: Comparison of several methods for estimating $N$ when the abundance distribuition is non convex: bias, standard-error, prediction error and non-coverage probabilities are reported, for $N = 100$ and $N = 5000$, considering $q_0 = 0.0707$ and $q_0 = 0.354$.

| | Emp | CvxPi | ChaoBunge | chao84 | ChaoLee | LB |
|---|---|---|---|---|---|---|
| | | | $N = 100, q_0 = 0.0707$ | | | |
| bias/N | -0.004 | -0.023 | 0.039 | 0.014 | 0.086 | -0.52 |
| se/N | 0.063 | 0.048 | 0.034 | 0.067 | 0.026 | 3.12 |
| EP/N | 0.063 | 0.049 | 0.051 | 0.069 | 0.09 | 3.16 |
| $P(B_{\inf} > N)$ | 0.4 | 0.2 | 0 | 2 | 45 | 0 |
| $P(B_{\sup} < N)$ | 7 | 3.7 | 38 | 6.7 | 38 | 23 |
| | | | $N = 100, q_0 = 0.354$ | | | |
| | Emp | CvxPi | ChaoBunge | chao84 | ChaoLee | LB |
| bias/N | -0.11 | -0.13 | -0.074 | 0.025 | 0.188 | -0.288 |
| se/N | 0.13 | 0.12 | 4.44 | 0.16 | 0.102 | 0.77 |
| EP/N | 0.18 | 0.17 | 4.44 | 0.16 | 0.21 | 0.82 |
| $P(B_{\inf} > N)$ | 9.2 | 9.3 | 0.7 | 1.4 | 96 | 0 |
| $P(B_{\sup} < N)$ | 0.4 | 0.6 | 3.3 | 3.4 | 2 | 12 |
| | | | $N = 5000, q_0 = 0.0707$ | | | |
| | Emp | CvxPi | ChaoBunge | chao84 | ChaoLee | LB |
| bias/N | -0.003 | -0.005 | 0.042 | 0.033 | 0.049 | 0.013 |
| se/N | 0.0109 | 0.007 | 0.004 | 0.006 | 0.004 | 0.015 |
| EP/N | 0.010 | 0.009 | 0.043 | 0.034 | 0.049 | 0.0189 |
| $P(B_{\inf} > N)$ | 5.4 | 5.0 | 0 | 0 | 0 | 0 |
| $P(B_{\sup} < N)$ | 1.7 | 1.23 | 100 | 100 | 100 | 26 |
| | | | $N = 5000, q_0 = 0.354$ | | | |
| | Emp | CvxPi | ChaoBunge | chao84 | ChaoLee | LB |
| bias/N | -0.11 | -0.11 | -0.11 | 0.055 | 0.027 | 0.055 |
| se/N | 0.019 | 0.019 | 0.04 | 0.019 | 0.023 | 0.055 |
| EP/N | 0.12 | 0.12 | 0.11 | 0.058 | 0.035 | 0.078 |
| $P(B_{\inf} > N)$ | 100 | 100 | 79 | 0 | 54 | 10 |
| $P(B_{\sup} < N)$ | 0 | 0 | 0 | 74 | 0.1 | 0 |

Let us first note that some of these methods require the choice of a cutoff value, denoted $t$, since only the less abundant species are used in the estimation procedures. The behavior of the algorithm as well as the estimations of $N$ may strongly depend on the choice of $t$. For the methods `unpmle`, `pnpmle` and `pcg`, we chose this $t$ according to the authors recommendations under the condition that the algorithm did converge, and to the goodness-of-fit of the empirical frequencies. For the `ChaoBunge` procedure we chose $t = 10$ according to the authors recommendations. In cases where the estimation of $N$ was negative, we decreased $t$ such that the resulting estimator was positive.

In the Microbial and Tomato datasets, where the empirical distribution is convex, it appears that the `unpmle` estimators and the `pcg` estimator (for the Tomato data) are almost equal to the empirical distribution, as it is obviously the case for our estimator. However the estimated values of $N$ differ a lot from a method to one another : from 4439 for our estimator, to 7417 for the `unpmle` estimator, up to 13960 for the `pcg` one.

In the Malayan Butterfly datasets, the empirical distribution is nearly convex for the less abundant species, and this is the case for all estimators. The estimated values of $N$ are less variable than for the two preceding examples, our method giving the highest value.

In the Bird example, the empirical distribution is non convex, and in particular, $f_1 < f_2$. Nevertheless, the estimator of $p^+$ based on the `pcg` method is convex. The estimators based on the two non-parametric Poisson mixture methods are non convex but $\widehat{p}_1^+$ and $\widehat{p}_2^+$ are far above $f_1$ and $f_2$. It is not easy to decide based on Figure 1 which estimate of $p^+$ should be preferred.

# 7 Proofs

## 7.1 Proof of Lemma 1

It follows from (2) that $D$ is distributed as a binomial variable with parameters $N$ and $1 - p_0$. Now, let $\{T_1, T_2, \ldots, T_D\}$ be the indices of the observed species. This means that we observe $A_{T_1}, \ldots, A_{T_D}$, and that $A_i = 0$ for all $i \notin \{T_1, T_2, \ldots, T_D\}$. The set $\{T_1, T_2, \ldots, T_D\}$ is random and takes values in $\mathcal{S}_{N,D}$, the set of all subsets of $\{1, \ldots, N\}$ with cardinality $D$. For all $d = 0, \ldots, N$ and all integers $a_i \geqslant 1$, $i = 1, \ldots d$, we have

$$P(A_{T_1} = a_1, \ldots, A_{T_D} = a_D | D = d)$$
$$= \frac{1}{P(D = d)} P(A_{T_1} = a_1, \ldots, A_{T_d} = a_d, \text{ and } A_i = 0 \text{ for all } i \notin \{T_1, \ldots, T_d\})$$
$$= \sum_{\{i_1, \ldots, i_d\} \in \mathcal{S}_{N,d}} \frac{1}{P(D = d)} P(A_{i_1} = a_1, \ldots, A_{i_d} = a_d, \text{ and } A_i = 0 \text{ for all } i \notin \{i_1, \ldots, i_d\}).$$

Table 3: Estimation of $N$ and 95% confidence intervals (between parenthesis). The values of the parameter $t$ are given in Figure 1 for the three last methods. For the `ChaoBunge` method, $t = 10$ for the Butterfly and Bird data sets, and $t = 3$ for the two others. The confidence interval could not be caculated in three of the four examples, because of convergence difficulties.

|  | Microbial | | Butterfly | | Bird | | Tomato | |
|---|---|---|---|---|---|---|---|---|
| `Emp` | 1211 | (1117, 1305) | 782 | (730, 834) | 82 | (66, 98) | 4439 | (4257, 4621) |
| `Cvx` | 1211 | (1117, 1305) | 782 | (730, 834) | 87 | (71, 96) | 4439 | (4257, 4621) |
| `ChaoBunge` | 2269* | (1213, 4821) | 757 | (698, 826) | 80 | (72, 92) | 7166* | (5330, 9947) |
| `ChaoLee` | 2511 | (1878, 3434) | 737 | (693, 787) | 80 | (72, 91) | 9554 | (7778, 11858) |
| `chao1984` | 1631 | (1326, 2050) | 714 | (679, 770) | 77 | (73, 92) | 5888 | (5275, 6610) |
| `LB` | 3987 | (915, 7060) | 754 | (629, 878) | 78 | (65, 93) | 11520 | (7047, 15993) |
| `pcg` | 3000 | NA | 744 | NA | 86 | (75, 95) | 13960 | NA |
| `pnpmle` | 2035 | (1523, 2758) | 724 | (686, 843) | 79 | (73, 100) | 7257 | (5899, 9167) |
| `upnpmle` | 2169 | $(1620, > 10^6)$ | 722 | (687, 913) | 76 | (74, 86) | 7417 | $(6009, > 10^7)$ |

The $A_i$'s are i.i.d. with distribution $p$, and $D$ is a binomial variable with parameters $N$ and $1 - p_0$, so denoting by $C_N^d$ the cardinality of $\mathcal{S}_{N,d}$, we obtain

$$\begin{aligned} P(A_{T_1} = a_1, \ldots, A_{T_D} = a_D | D = d) &= \sum_{\{i_1,\ldots,i_d\} \in \mathcal{S}_{N,d}} \frac{p_{a_1} \times \cdots \times p_{a_d} \times p_0^{N-d}}{C_N^d (1-p_0)^d p_0^{N-d}} \\ &= \sum_{\{i_1,\ldots,i_d\} \in \mathcal{S}_{N,d}} \frac{p_{a_1}^+ \times \cdots \times p_{a_d}^+}{C_N^d} \\ &= p_{a_1}^+ \times \cdots \times p_{a_d}^+. \end{aligned}$$

This proves that conditionally on $D$, the observations $A_{T_1}, \ldots, A_{T_D}$ are i.i.d with distribution $p^+$. Setting $X_i = A_{T_i}$ for all $i = 1, \ldots, D$ completes the proof of the lemma. □

## 7.2 Proof of (15)

Thanks to (8), we have $\mathbb{E}(\widehat{N}^f) = N$. Moreover,

$$\begin{aligned} \widehat{N}^f &= 2S_1 - S_2 + S_1 + S_2 + S_3 + \ldots \\ &= 3S_1 + \sum_{j \geqslant 3} S_j, \end{aligned}$$

so that

$$\mathbb{V}(\widehat{N}^f) = 9Np_1(1-p_1) + Np_{\geqslant 3}(1-p_{\geqslant 3}) - 6Np_1 p_{\geqslant 3} = 6Np_1,$$

where $p_{\geqslant 3} = \sum_{j \geqslant 3} p_j = 1 - 3p_1$. From the central limit theorem, it follows that

$$\frac{\widehat{N}^f - N}{\sqrt{N}} \text{ converges in law to } \mathcal{N}(0, 6p_1), \tag{22}$$

as $N \to \infty$. Since $S_1$ is distributed as a binomial variable with parameters $N$ and $p_1$, this yields (15). $\square$

## 7.3 Proof of Theorem 1

For the sake of simplicity the function $\Phi_{\mathcal{I}}$ will be denoted by $\Phi$.

Consider the random vectors $Y$, $Z$ and $U$ defined as follows:

$$Y = \sqrt{D}\left(\widehat{p}^{+,k} - p^{+,k}\right), \; Z = \sqrt{D}\left(\Phi(f^k) - p^{+,k}\right), \; U = \sqrt{D}\left(f^k - p^{+,k}\right)$$

where $p^{+,k} = (p_1^+, \ldots, p_k^+)$, $\widehat{p}^{+,k} = (\widehat{p}_1^+, \ldots, \widehat{p}_k^+)$, and $f^k = (f_1, \ldots, f_k)$, with $f_l = S_l/D$ denoting the empirical estimator of $p_l^+$ for all $l = 1, \ldots, k$. In order to prove the theorem, we will prove the following assertions:

(i). The random vector $Z$ satisfies $Z = \Phi(U)$.

(ii). The function $\Phi$ is continuous.

(iii). The probability that $(Z_1, Z_2) = (Y_1, Y_2)$ tends to one as $N \to \infty$.

(iv). The random vector $(U, V_{k+1})$ converges in distribution to $(W, T_0)$ as $N \to \infty$, where $V_{k+1} = (D\theta - N)/\sqrt{N(\theta - 1)}$ and $T_0$ s a standard Gaussian variable independent of $W$.

By continuity of $\Phi$, we derive from (iv) that $(\Phi(U), V_{k+1})$ converges in law to $(\Phi(W), T_0)$. Thanks to (i), this implies that $(Z, V_{k+1})$ converges in law to $(\Phi(W), T_0)$. We then conclude from (iii) that $(Y_1, Y_2, V_{k+1})$ converges in law to $(\Phi(W)_1, \Phi(W)_2, T_0)$. In particular, the variable $\sqrt{D}(\widehat{\theta} - \theta) = 2Y_1 - Y_2$ converges in distribution to $2\Phi(W)_1 - \Phi(W)_2$, which proves the first assertion. Moreover, we have

$$\begin{aligned}
\frac{\widehat{N} - N}{\sqrt{D}} &= (1 + o_P(1))\frac{D\widehat{\theta} - N}{\sqrt{D}} \\
&= (1 + o_P(1))\frac{D\theta - N}{\sqrt{D}} + (1 + o_P(1))\sqrt{D}(\widehat{\theta} - \theta) \\
&= (1 + o_P(1))\frac{D\theta - N}{\sqrt{N(1 - p_0)}} + (1 + o_P(1))\sqrt{D}(\widehat{\theta} - \theta),
\end{aligned}$$

since $D$ has a binomial distribution with parameters $N$ and $1 - p_0$ (see Lemma 1). We conclude that $(\widehat{N} - N)/\sqrt{D}$ converges in distribution to $2\Phi(W)_1 - \Phi(W)_2 + T$, where $T$ is a centered Gaussian variable with variance $(\theta - 1)/(1 - p_0) = \theta(\theta - 1)$, independent of $2\Phi(W)_1 - \Phi(W)_2$. This concludes the proof of Theorem 1. $\square$

**Proof of (i)** Both $q - \sqrt{D}p^{+,k}$ and $q + \sqrt{D}p^{+,k}$ belong to $\mathcal{C}^{\mathcal{I}}$ for all $q \in \mathcal{C}^{\mathcal{I}}$ since $p^{+,k}$ is linear on each interval $[i_{j-1}, i_j]$ for $j = 1, \ldots, I$. Therefore we have

$$\Phi(U) = \arg\min_{q \in \mathcal{C}^{\mathcal{I}}} \sum_{l=1}^{k} \left(q_l + \sqrt{D}p_l^+ - \sqrt{D}f_l\right)^2$$

$$= \arg\min_{q \in \mathcal{C}^{\mathcal{I}}} \sum_{l=1}^{k} \left(q_l - \sqrt{D}f_l\right)^2 - \sqrt{D}p^{+,k}.$$

Therefore, $\Phi(U) = \sqrt{D}\Phi(f^k) - \sqrt{D}p^{+,k} = Z$.

**Proof of (2)** By definition, $\Phi$ is the projection operator from $\mathbb{R}^k$ to the closed convex subset $\mathcal{C}^{\mathcal{I}}$ of $\mathbb{R}^k$, so

$$\sum_{l=1}^{k} (\Phi(t)_l - \Phi(u)_l)^2 \leqslant \sum_{l=1}^{k} (t_l - u_l)^2$$

for all $t, u \in \mathbb{R}^k$ (see (3) in Durot et al. (2013) for similar arguments). Therefore, the left hand side tends to zero as soon as the right hand side tends to zero, which proves that $\Phi$ is continuous.

**Proof of (3)** Using notation of Section 2.1, we can write

$$f_l = \frac{1}{D} \sum_{i=1}^{N} I(A_i = l),$$

where the $A_i$'s are i.i.d. with distribution $p$, where $D = \sum_{i=1}^{N} I(A_i > 0)$ is a binomial variable with parameters $N$ and $1 - p_0$. The variable $D/N$ converges in probability to $1 - p_0$ as $N \to \infty$, so we have

$$f_l = (1 + o_P(1)) \frac{1}{(1 - p_0)N} \sum_{i=1}^{N} I(A_i = l).$$

From the law of large numbers, we conclude that

$$f_l \text{ converges in probability to } \frac{p_l}{1 - p_0} = p_l^+ \qquad (23)$$

for all $l \geqslant 1$, as $N \to \infty$.

Now, let us define $\widehat{q}$ as follows: denoting by $k_0$ the $\min\{\tau + 1, \kappa\}$ so that $k \geqslant k_0$, let $\widehat{q}_l = \Phi(f^k)_l$ for all $l \in \{1, \ldots, k_0\}$ and let $\widehat{q}_l = \widehat{p}_l^+$ for all $l \geqslant k_0 + 1$. From the definition of $\Phi$ together with the convexity of $\widehat{p}^+$, it follows that $\widehat{q}$ is piecewise convex and the only points of non-convexity, if any, are $k_0$, $k_0 + 1$ and the points in $\mathcal{I}$ that are smaller than $k_0$. This means that the set $\mathcal{I}_{\widehat{q}}$ of non-convexity points of $\widehat{q}$ can only contains $k_0$, $k_0 + 1$ and knots of $p^+$ that are smaller than $k_0$. We will show that $\widehat{q}$ is convex with probability that tends to one, which amounts to prove that for all $l \in \mathcal{I}_{\widehat{q}}$, the probability $P(\Delta(\widehat{q}_l) \geqslant 0)$ tends to one as $N$ goes to infinity, where we set

$$\Delta(q_l) = q_{l+1} - 2q_l + q_{l-1}$$

for all sequences $q = (q_1, q_2, \ldots)$. Because $p^{+,k}$ belongs to $\mathcal{C}_\mathcal{I}$, it follows from the definition of $\Phi(f^k)$ that
$$\sum_{l=1}^{k} \left(\Phi(f^k)_l - f_l\right)^2 \leqslant \sum_{l=1}^{k} \left(p_l^+ - f_l\right)^2.$$
Combining this with (23) proves that $\Phi(f^k)_l - f_l$ converges in probability to 0 for all $l = 1, \ldots, k$, which means that $\Phi(f^k)_l$ converges in probability to $p_l^+$. A similar argument proves that $\widehat{p}_{k_0+1}^+$ and $\widehat{p}_{k_0+2}^+$ converge in probability to $p_{k_0+1}^+$ and $p_{k_0+2}^+$ respectively, so we conclude that for all $l \geqslant 1$, the variable $\Delta(\widehat{q}_l)$ converges to $\Delta(p_l^+)$ in probability as $N \to \infty$. Let us notice that in the case where $k_0 = \tau + 1$, the assumption that $k_0$ is finite ensures that $p_0$ has a finite support, and $\Delta p_{\tau+1}^+ = p_\tau^+ > 0$, whereas in the case where $k_0 = \kappa$, $k_0$ is clearly a knot of $p^+$. Therefore, $\Delta(p_l^+) > 0$, for all $l \in \mathcal{I}_{\widehat{q}} \setminus \{k_0 + 1\}$. Using that $\Delta(\widehat{q}_l) - \Delta(p_l^+)$ tends to 0 in probability, we conclude that
$$P\left(\Delta(\widehat{q}_l) < 0\right) \leqslant P\left(\Delta(\widehat{q}_l) - \Delta(p_l^+) < -\Delta(p_l^+)/2\right)$$
where the right-hand side tends to 0 for all $l \in \mathcal{I}_{\widehat{q}} \setminus \{k_0+1\}$. This shows that $P\left(\Delta(\widehat{q}_l) \geqslant 0\right)$ tends to 1 for all $l \in \mathcal{I}_{\widehat{q}} \setminus \{k_0 + 1\}$. It remains to prove that with probability tending to one, $\widehat{q}$ is convex at $k_0 + 1$. To do this, we consider two cases.

- If $k_0 < \tau + 1$, then $k_0$ is a double knot of $p^+$, so that $\Delta(p_{k_0+1}^+) > 0$. Following the same argument as before, we conclude that $P\left(\Delta(\widehat{q}_{k_0+1}) \geqslant 0\right)$ tends to one.

- If $k_0 = \tau + 1$, then $\Delta(\widehat{q}_{k_0+1}) = \widehat{p}_{k_0+2}^+ - 2\widehat{p}_{k_0+1}^+ + \widehat{q}_{k_0}$ converges in probability to $\Delta(p_{k_0+1}^+) = 0$ so the above arguments do not apply. In that case, the support of $p^+$ is finite and with probability that tends to one, the maximum of the support of $\widehat{p}_n^+$ is either $\tau$ or $\tau + 1$, see Balabdaoui et al. (2014). This implies that $\Delta(\widehat{q}_{k_0+1}) = \widehat{q}_{k_0}$ with probability that tends to one. It can be proved that $\widehat{q}_{k_0} \geqslant 0$ (see the proof of Lemma 1 in Durot et al. (2013) for similar arguments), so $P\left(\Delta(\widehat{q}_{k_0+1}) \geqslant 0\right)$ tends to one.

Now that we have proved that $\widehat{q}$ is convex with probability that tends to one, note that from the definition of $\widehat{p}^+$, it follows that with probability tending to one,
$$\sum_{l \geqslant 1} \left(\widehat{p}_l^+ - f_l\right)^2 \leqslant \sum_{l \geqslant 1} \left(\widehat{q}_l - f_l\right)^2,$$
or equivalently,
$$\sum_{l=1}^{k_0} \left(\widehat{p}_l^+ - f_l\right)^2 \leqslant \sum_{l=1}^{k_0} \left(\Phi(f^k)_l - f_l\right)^2. \tag{24}$$

Now, let us define $\widehat{r}$ as follows: denoting by $\mathcal{I}_0$ the set of all points in $\mathcal{I}$ that are smaller than or equal to $k_0$, let $\widehat{r}_l = \Phi_{\mathcal{I}_0}(f^{k_0})_l$ for all $l \in \{1, \ldots, k_0\}$ and let $\widehat{r}_l = \Phi(f^k)_l$ for all $l$ with $l \geqslant k_0 + 1$ and $l \leqslant k$ (note that $\widehat{r} = \Phi(f^k)$ in the particular case where $k = k_0$). Similar arguments as above prove that with probability that tends to one, $\widehat{r} \in \mathcal{I}$ so that
$$\sum_{l=1}^{k} \left(\Phi(f^k)_l - f_l\right)^2 \leqslant \sum_{l=1}^{k} \left(\widehat{r}_l - f_l\right)^2$$

or equivalently,
$$\sum_{l=1}^{k_0} \left(\Phi(f^k)_l - f_l\right)^2 \leqslant \sum_{l=1}^{k_0} \left(\Phi_{\mathcal{I}_0}(f^{k_0})_l - f_l\right)^2.$$

But, $(\Phi(f^k)_1, \ldots, \Phi(f^k)_{k_0}) \in \mathcal{C}^{\mathcal{I}_0}$ so it follows from the definition of $\Phi_{\mathcal{I}_0}$ that
$$\sum_{l=1}^{k_0} \left(\Phi_{\mathcal{I}_0}(f^{k_0})_l - f_l\right)^2 \leqslant \sum_{l=1}^{k_0} \left(\Phi(f^k)_l - f_l\right)^2.$$

Therefore, we have an equality with probability that tends to one, and combining this with (24) yields
$$\sum_{l=1}^{k_0} \left(\Phi_{\mathcal{I}_0}(f^{k_0})_l - f_l\right)^2 = \sum_{l=1}^{k_0} \left(\Phi(f^k)_l - f_l\right)^2$$
$$\geqslant \sum_{l=1}^{k_0} \left(\widehat{p}_l^+ - f_l\right)^2.$$

By convexity of $\widehat{p}^+$ we have $(\widehat{p}_1^+, \ldots, \widehat{p}_{k_0}^+) \in \mathcal{C}^{\mathcal{I}_0}$, so we also have
$$\sum_{l=1}^{k_0} \left(\Phi_{\mathcal{I}_0}(f^{k_0})_l - f_l\right)^2 \leqslant \sum_{l=1}^{k_0} \left(\widehat{p}_l^+ - f_l\right)^2,$$

so again, we have in fact an equality with probability tending to one. This implies that
$$\sum_{l=1}^{k_0} \left(\Phi_{\mathcal{I}_0}(f^{k_0})_l - f_l\right)^2 = \sum_{l=1}^{k_0} \left(\Phi(f^k)_l - f_l\right)^2 = \sum_{l=1}^{k_0} \left(\widehat{p}_l^+ - f_l\right)^2,$$

with probability tending to one. Since $\Phi_{\mathcal{I}_0}(f^{k_0})$ is uniquely defined, this proves that $\Phi_{\mathcal{I}_0}(f^{k_0})_l = \Phi(f^k)_l = \widehat{p}_l^+$ for all $l \in \{1, \ldots, k_0\}$ with probability tending to one. This completes the proof of (iii) since $k_0 \geqslant 2$. $\square$

**Proof of (iv)** Using the notation of Section 2.1, we can write
$$U_l = \frac{1}{\sqrt{D}} \sum_{i=1}^{N} \left(I(A_i = l) - p_l^+ I(A_i > 0)\right).$$

But $D/N$ converges in probability to $1 - p_0 = 1/\theta$, so
$$U_l = (1 + o_P(1))\frac{1}{\sqrt{N}} \sum_{i=1}^{N} \sqrt{\theta} \left(I(A_i = l) - p_l^+ I(A_i > 0)\right).$$

Here, the random variables $\sqrt{\theta}(I(A_i = l) - p_l^+ I(A_i > 0))$ are i.i.d. with mean
$$\sqrt{\theta}(p_l - p_l^+(1 - p_0)) = 0$$

and variance

$$\theta\Big(p_l(1-p_l) + (p_l^+)^2 p_0(1-p_0) - 2p_l^+ p_l p_0\Big) = \theta\Big(p_l(1-p_l) - (p_l^+)^2 p_0(1-p_0)\Big)$$
$$= \theta\Big(p_l - (p_l^+)^2(1-p_0)\Big)$$
$$= p_l^+(1-p_l^+).$$

Besides,

$$V_{k+1} = \frac{D\theta - N}{\sqrt{N(\theta-1)}} = \frac{1}{\sqrt{N(\theta-1)}} \sum_{i=1}^{N} (\theta I(A_i > 0) - 1),$$

where the variables $\theta I(A_i > 0) - 1$ are i.i.d. with mean zero and variance $\theta p_0 = \theta - 1$. Dealing simultaneously with all components of $(U, V_{k+1})$, we derive from the vectorial central limit theorem that $(U, V_{k+1})$ converges in distribution as $N \to \infty$ to a Gaussian random vector with mean zero and variance matrix $\widetilde{\Gamma}$ with component $(l, j)$ defined as follows: for all $j = l$,

$$\widetilde{\Gamma}_{l,l} = \begin{cases} p_l^+(1-p_l^+), & \text{for } l = 1, \ldots k \\ 1, & \text{for } l = k+1, \end{cases}$$

and for all $j < l$,

$$\widetilde{\Gamma}_{l,j} = \begin{cases} \theta cov(I(A_i = l) - p_l^+ I(A_i > 0), I(A_i = j) - p_j^+ I(A_i > 0)), & \text{for } l \leqslant k \\ \sqrt{\theta/(1-\theta)} cov(I(A_i = j) - p_j^+ I(A_i > 0), \theta I(A_i > 0) - 1), & \text{for } l = k+1 \end{cases}$$

$$= \begin{cases} \theta E\Big[\Big(I(A_i = l) - p_l^+ I(A_i > 0)\Big)\Big(I(A_i = j) - p_j^+ I(A_i > 0)\Big)\Big], & \text{for } l \leqslant k \\ \sqrt{\theta/(1-\theta)} E\Big[\Big(I(A_i = j) - p_j^+ I(A_i > 0)\Big)\theta I(A_i > 0)\Big], & \text{for } l = k+1 \end{cases}$$

$$= \begin{cases} \theta E\Big[-p_j^+ I(A_i = l) - p_l^+ I(A_i = j) + p_j^+ p_l^+ I(A_i > 0)\Big], & \text{for } l \leqslant k \\ \theta\sqrt{\theta/(1-\theta)} E\Big[I(A_i = j) - p_j^+ I(A_i > 0)\Big], & \text{for } l = k+1 \end{cases}$$

$$= \begin{cases} -p_j^+ p_l^+, & \text{for } l \leqslant k \\ 0, & \text{for } l = k+1. \end{cases}$$

Therefore,

$$\widetilde{\Gamma} = \begin{pmatrix} \Gamma & 0 \\ {}^t 0 & 1 \end{pmatrix},$$

where 0 denotes the null vector in $\mathbb{R}^k$. It follows from the assumptions on $W$ and $T_0$ that $(W, T_0)$ is a centered Gaussian vector in $\mathbb{R}^{k+1}$ with variance matrix $\widetilde{\Gamma}$. From what precedes, $(U, V_{k+1})$ thus converges in distribution to $(W, T_0)$. This concludes the proof of (iv). □

### 7.4 Some characteristics on the Gamma-Poisson distribution

**Lemma 2** *Let $p$ be the Gamma-Poisson distribution defined for all $j \geqslant 0$ by*

$$p_j = \frac{\Gamma(j+\nu)}{\Gamma(\nu) j!} \mu^\nu (1-\mu)^j,$$

*for some $\nu > 0$ and $\mu \in (0, 1)$. Then $p$ is convex if and only if either $0 < \nu < 1$, or $\nu \geqslant 1$ and $1 - \mu \leqslant r$, where*

$$r = \frac{2\nu - \sqrt{2\nu(\nu - 1)}}{\nu(\nu + 1)}.$$

*Moreover, $p$ is a convex abundance distribution if $\nu \geqslant 1$ and $1 - \mu = r$.*

**Proof** A distribution $p$ is a convex abundance distribution if $\pi_j \geqslant 0$ for all integers $j \geqslant 2$ and $\pi_1 = 0$, where $\pi_j$ is given by Equation (7). Using that $\Gamma(a+1) = a\Gamma(a)$ for all $a > 0$, one obtains that $\pi_j \geqslant 0$ if and only if

$$(j + \nu)(j + \nu - 1)(1 - \mu)^2 - 2(j + \nu - 1)(j + 1)(1 - \mu) + j(j + 1) \geqslant 0.$$

By standard calculation we obtain that if $\nu < 1$, then $\pi_j > 0$ for all $j$. Let $\nu \geqslant 1$ and

$$r(j, \nu) = \frac{(j+1)(j+\nu-1) - \sqrt{(j+1)(j+\nu-1)(\nu-1)}}{(j+\nu-1)(j+\nu)}$$

$$s(j, \nu) = \frac{(j+1)(j+\nu-1) + \sqrt{(j+1)(j+\nu-1)(\nu-1)}}{(j+\nu-1)(j+\nu)}$$

It is easy to verify that for all $j \geqslant 1$, $0 < r(j, \nu) \leqslant 1$ for all $\nu \geqslant 1$ and $s(j, \nu) < 1$ for all $\nu > 2$.

Therefore $\pi_j \geqslant 0$ if and only if

$$(\nu, 1 - \mu) \in \{(0, 1] \times (0, 1)\} \cup \{[1, \infty] \times (0, r(j, \nu)]\} \cup \{(2, \infty) \times [s(j, \nu), 1)\}.$$

It follows that $p$ is convex for all $\mu \in (0, 1)$ and $0 < \nu \leqslant 1$. Let us now study the case where $\nu > 1$: $p$ will be convex if for all $\nu > 1$, $1 - \mu \leqslant \inf_{j \geqslant 1} r(j, \nu)$, or if for all $\nu \geqslant 2$, $1 - \mu \geqslant \sup_{j \geqslant 1} s(j, \nu)$.

The second condition cannot be satisfied since the sequence $s(j, \nu)$ tends to 1 when $j$ tends to infinity.

Let us now show that $r(j, \nu) > r(1, \nu)$ for all $j \geqslant 2$. Calculating the derivative of the function $r$ with respect to $j$, we obtain

$$\frac{(j+\nu)^2}{\sqrt{\nu-1}} r'(j, \nu) = \sqrt{\nu-1} + \sqrt{\frac{1}{j+\nu-1}} \left( \sqrt{j+1} - \frac{(j+\nu)(\nu-2)}{2(j+\nu-1)\sqrt{j+1}} \right).$$

Therefore $r'(j, \nu) \geqslant 0$ if and only if

$$\underbrace{2(j+1)(j+\nu-1) - (\nu-2)(j+\nu)}_{P_1(j,\nu)} \geqslant -\underbrace{2(j+\nu-1)^{3/2}\sqrt{(j+1)(\nu-1)}}_{P_2(j,\nu)}.$$

Let $\nu_j = (j+4+\sqrt{9j^2+16j+8})/2$. If $1 \leqslant \nu \leqslant \nu_j$, then $P_1(j, \nu) \geqslant 0$ and thus $r'(j, \nu) \geqslant 0$. It remains to consider the case where $\nu > \nu_j$. When $j \geqslant 2$,

$$\frac{P_2(j, \nu)}{2\sqrt{j+1}} \geqslant (\nu+1)^{3/2}\sqrt{\nu-1} \geqslant \nu^2 \text{ as soon as } \nu \geqslant 2$$

and
$$-P_1(j,\nu) = \nu^2 - (j+4)\nu - 2(j^2 + j - 1) \leqslant \nu^2.$$

Therefore $-P_1(j,\nu) \leqslant P_2(j,\nu)$ if $\nu > \nu_j$. Finally we have shown that $r(j,\nu) \geqslant r(1,\nu)$ for all $j \geqslant 1$ and $\nu > 1$, which proves that $p$ is convex for all $\nu > 1$ and

$$0 < 1 - \mu \leqslant r(1) = \frac{2\nu - \sqrt{2\nu(\nu-1)}}{\nu(\nu+1)}.$$

# Acknowledgements